\documentclass[aps,prl,reprint,showpacs,amsfonts,amsmath]{revtex4-1}
\usepackage[final]{graphicx}

\newcommand\sm{\text{s}}
\newcommand\la{\text{l}}

\begin{document}
\title{Multiple Glasses in Asymmetric Binary Hard Spheres}
\date{\today}
\newcommand\dlr{\affiliation{Institut f\"ur Materialphysik im Weltraum,
  Deutsches Zentrum f\"ur Luft- und Raumfahrt (DLR), 51170 K\"oln,
  Germany}}
\newcommand\ukn{\affiliation{Fachbereich Physik, Universit\"at Konstanz,
  78457 Konstanz, Germany}}
\newcommand\zkk{\affiliation{Zukunftskolleg, Universit\"at Konstanz,
  78457 Konstanz, Germany}}
\author{Th.~Voigtmann}\dlr\ukn\zkk

\begin{abstract}
Multiple distinct glass states occur in binary hard-sphere mixtures
with constituents of very disparate sizes according to the mode-coupling
theory of the glass transition (MCT),
distinguished by considering
whether small particles remain mobile or not, and whether small particles
contribute significantly to perturb the big-particle structure or not. In
the idealized glass, the four different glasses are separated by sharp
transitions that give rise to higher-order transition phenomena
involving logarithmic decay laws, and to
anomalous power-law-like diffusion. The phenomena are argued to be
expected generally in glass-forming mixtures.
\end{abstract}

\pacs{pacs tbd}

\maketitle

Many glass formers and virtually
all simple model systems for slow dynamics are mixtures of some sort.
Close to a glass transition generic
mixing effects appear, such as dramatic changes in viscosity induced by
small composition changes \cite{Williams2001b,Foffi2003},
that are relevant for applications and may
help to shed light on the microscopic processes driving
glass formation.

Even more interesting is the possibility to form
qualitatively distinct types of glass, depending on mixture composition
and constituents.
Taking for example binary mixtures with sufficiently disparate constituents,
a glass can form where some (slow) species freeze, but a fast component
is able to diffuse through the voids left in the amorphous packing.
This scenario is particularly relevant for transport through heterogeneous
disordered media \cite{Krako2005,Krako2006,Kahl,Kim,Hoefling.2006} or glassy
ion conductors \cite{Voigtmann2006,wcamix}. The simplest model are binary
hard-sphere
mixtures with large size disparity, where experiments on colloidal suspensions
indeed found, depending on relative concentration, a partially frozen
``single glass'' with mobile small particles, separate from a ``double glass''
where both particle species freeze \cite{Imhof1995,Imhof1995b}.
In mixtures of star polymers
\cite{Zaccarelli2005,Mayer2007,Mayer.2008} yet
another kind of glass emerged,
termed ``asymmetric'' because it is characterized by few big particles frozen
in a small-particle matrix, rendering the big-particle nearest-neighbor
cages highly nonspherical.
It was, however, argued to be a hallmark of the ultra-soft
interactions typical for the star polymers.

Another kind of glass intuitively argued for is the ``attractive glass''
famous from colloid--polymer mixtures where free polymer induces
depletion attraction among the colloids.
If that attraction is weak, the glass that forms is essentially
hard-sphere like or ``repulsive'', while at sufficiently strong
and short-ranged attraction, a new glass driven by bonding and not
nearest-neighbor cageing appears. Based on extensively tested predictions
of the mode-coupling theory of the glass transition (MCT) for a square-well
model system
\cite{Goetze2009,Dawson2001,Pham2002,Pham2004,
Sciortino2003,Zaccarelli2003b}, one expects the two glasses to be
separated by a glass--glass transition
crossing which, for example, the elastic moduli of the glass
exhibit sharp changes.
Considering the generality of the depletion-interaction mechanism
\cite{Asakura1958}, one may indeed expect a similar glass--glass transition
to be present in binary mixtures quite generically (lest the relevant parameter
space cannot be explored).
As such transitions typically involve endpoint
singularities that give rise to universal logarithmic decay laws for the
time-dependent correlation functions \cite{Sperl2003a},
one anticipates regions of mixture composition where these peculiar decay laws
can be found. Indeed, they have been reported in computer simulation
of soft-sphere mixtures \cite{Moreno2006b}.

Here I demonstrate, that already the simplest glass-forming
mixture model, the binary hard-sphere mixture,
allows to identify \emph{four} qualitatively different
glassy states separated by well-defined transitions.
The transition diagram lends itself to an intuitive classification:
one has to distinguish (i) whether small particles remain mobile in the
glass or not, and (ii) whether the structure of the big-particle glass
is or is not significantly disturbed by the small particles. Combining
these two choices each gives four possible types of glass.

Calculations are based on MCT supplemented
by the Percus-Yevick (PY) approximation for the static structure factor
\cite{Goetze1987b,Goetze2003,Hajnal2009}, but following the physically plausible
classification, the results can be expected to hold rather generally for
binary mixtures whose parameters can be tuned widely enough.
The possible interplay with equilibrium phases is ignored here.
The resulting glass-transition diagram is a unique
prediction of MCT, distinct from other theories that have been put forward
\cite{JuarezMaldonado2008},
and testable in simulation or experiment.

Numerical calculations follow the method of Ref.~\cite{Goetze2003}.
MCT takes partial static structure factors
$S_{\alpha\beta}(q)=\langle\varrho_\alpha^*(\vec q)\varrho_\beta(\vec q)
\rangle$ as input (Greek indices label species, and static triplet
correlations are neglected here).
Here $\langle\cdot\rangle$ is the canonical
average, and $\varrho_\alpha(\vec q)=\sum_k\exp[i\vec q\vec r_{k,\alpha}]$
is the number-density fluctuation of species $\alpha$ (the sum runs over all
particle positions $\vec r$ of that type).
MCT predicts collective dynamical density correlation functions
$\Phi_{\alpha\beta}(q,t)=\langle\varrho_\alpha^*(\vec q,t)\varrho_\beta(\vec q)
\rangle$, where the time evolution is given by the implicit dependence of
$\varrho_\alpha(\vec q)$ on the trajectories.
The long-time limit of
that matrix, $\boldsymbol F(q)=\lim_{t\to\infty}\boldsymbol\Phi(q,t)$,
distinguishes ergodic liquid states, $\boldsymbol F(q)=\boldsymbol0$,
from nonergodic ones, where the correlation matrix decays not to zero but
to a positive definite matrix, $\boldsymbol F(q)\succ\boldsymbol0$.
Such states are identified as (idealized) glasses within MCT.
Standard MCT glass transitions are characterized as bifurcations of the
${\mathcal A}_2$ type where $\boldsymbol F(q)$ jumps discontinuously.
Note that such jumps can also occur
inside the glass.
The $\boldsymbol F(q)$ can be
found numerically \cite{Franosch2002}
on a finite wave-number grid, $q_i=(i+1/2)\Delta q$
with $i=1,\ldots N=500$ and $\Delta q=0.4$ for most calculations here.
Small size ratios require higher large-$q$ cutoff, $N=1000$ was hence used
for $\delta<0.2$.

\begin{figure}
\includegraphics[width=.9\linewidth]{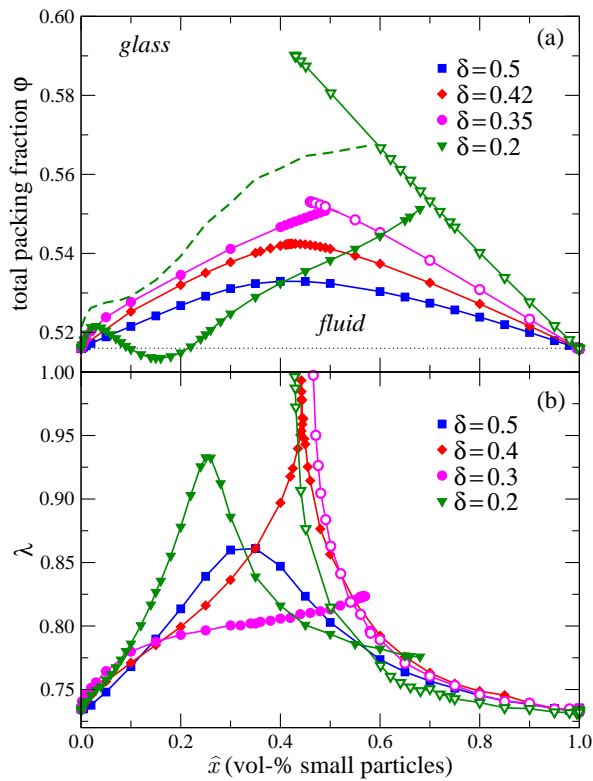}
\caption{\label{phicdiag}
  (a)
  Glass transition diagram for a model binary hard-sphere mixture obtained from
  mode-coupling theory. Solid lines show the glass-transition packing
  fraction $\varphi_c(\delta,\hat x)$ as a function of small-particle
  volume concentration, $\hat x$, for various size ratios $\delta$ as
  indicated. The dashed line indicates a separate localization transition
  $\varphi_c^s(\delta,\hat x)$
  for the small species for $\delta=0.2$.
  (b)
  MCT exponent parameter $\lambda(\hat x)$ for cuts of constant $\delta$.
  Solid and open symbols refer to the small-$\hat x$ and large-$\hat x$
  branches, respectively.
}
\end{figure}

Results are reported as cuts through the parameter space at
constant size ratio $\delta=d_\sm/d_\la\le1$, where
$d_\alpha$ are the hard-sphere diameters ($\alpha=\sm,\la$ for small and large).
The two remaining
parameters then are the number densities $\rho_\alpha$,
conveniently expressed as a
total packing fraction of the spheres, $\varphi=\sum_\alpha\varphi_\alpha
=(\pi/6)\sum_\alpha\rho_\alpha d_\alpha^3$, and a concentration (by volume)
of small spheres, $\hat x=\varphi_\sm/\varphi$.
To top panel of
Fig.~\ref{phicdiag} illustrates the glass-transition diagram for
size ratios $0.2\le\delta\le0.5$; larger $\delta$ are topologically
the same as $\delta=0.5$ \cite{Goetze2003}. For all compositions at fixed
$\delta\gtrsim0.42$, the binary-mixture glass is separated from the
liquid by a smooth line of ordinary ${\mathcal A}_2$ transitions.
In contrast, the curve for, say, $\delta=0.35$ demonstrates
the emergence of a glass--glass transition: the liquid-glass transition
splits into two lines that no longer
join smoothly, but cross discontinuously at some point, at which one of
the lines stops. The other continues inside the glass until it
terminates (the corresponding jump in $\boldsymbol F(q)$ disappears)
at a higher-order transition point of type ${\mathcal A}_3$.
The glass--glass transition signals that
there are two competing arrest mechanisms at work: cageing can either be
dominated by large particles or by small ones.
As soon as the ratio of
relevant length scales (essentially $\delta$) becomes sufficiently
distinct from unity, the two mechanisms give rise to glasses with differing
intrinsic scales (the localization length, or typical cage size) such that
a sharp distinction becomes possible. As pointed out, one way to
distinguish these glasses experimentally would be a marked difference in
elastic properties.
The structure of the
large-$\hat x$ glass bears resemblance to certain types of sweets
\cite{JuarezMaldonado2008} such as
Italian torroncino. The distinction is only strict for size
ratios $\delta<\delta_c$. At $\delta_c$,
the glass--glass transition endpoint coincides with the crossing,
resulting in an ${\mathcal A}_4$ singularity.
Mathematically, these
higher-order singularities are identical to those in
one-component descriptions of colloid-polymer mixtures
\cite{Sperl2003a}, and the same kind of asymptotic
expansions predicting the
the appearance of logarithmic decay laws and their precursors apply
here, too. In the present calculations, $\delta_c\approx0.41$.
Since the states $\hat x=0$ and $\hat x=1$ are identical up to a
rescaling of lengths with $\delta$, the $\boldsymbol F$-versus-$q$ curve
is broader for $\hat x\to1$ than for $\hat x\to0$, and generically
higher at fixed $q$. Hence the large-$\hat x$ portion of the transition
extends into and demarks the mechanically stiffer glass.


The position of higher-order singularities is revealed by looking at
MCT's exponent parameter $\lambda$. To each transition point, a particular
value of $0.5\le\lambda\le1$ is assigned in the theory, which determines
the power-law exponents valid for the asymptotic expansion of correlation
functions close to the transition. At $\lambda=1$, these power laws cease
to be valid and are replaced by
logarithmic laws. As seen in the lower panel of
Fig.~\ref{phicdiag}, the values of $\lambda$ indeed rise to unity
when approaching the endpoints of the large-$\hat x$ glass-transition lines.
At $\delta>\delta_c$, $\lambda$ still rises
as a precursor corresponding to a notable decrease in the asymptotic
exponents; at $\delta=\delta_c$, the
$\lambda(\hat x)$-curve reaches unity for the first time before splitting
into two branches.

MCT predicts that at all the ${\mathcal A}_\ell$ transition points
discussed so far, all collective partial-density
correlation functions show a simultaneous jump in their long-time limits
and hence all are nonergodic \cite{Franosch2002}.
However, this does not need to hold for the nonergodicity factors
$f^s_\alpha(q)=\lim_{t\to\infty}\phi^s_\alpha(q,t)$ associated with the
self-motion of the species $\alpha$,
$\phi^s_\alpha(q,t)
=\langle\exp[i\vec q(\vec r_{s,\alpha}(t)-\vec r_{s,\alpha}(0))]\rangle$.
$f^s_\sm(q)$
can remain zero when crossing the glass transition line, if the small
particles are below a certain size $\delta^s_c$
\cite{Sjoegren1986,Bosse1987}.
A dashed line in Fig.~\ref{phicdiag}(a) demarcs for $\delta=0.2$
the region where $f^s_\sm(q)=0$ and hence the ``single glass''.
The collective small--small density correlation
function remains nonergodic also in the single glass,
in distinct disagreement with
the SCGLE theory of Ref.~\cite{JuarezMaldonado2008}.

The nature of the transition lines shown in dashed in Fig.~\ref{phicdiag}
is notably different from the ${\mathcal A}_\ell$ glass transition lines shown
as solid lines: they are so-called type~A or localization transitions, where
the nonergodicity factor $f^s_{\sm}(q)$ rises continually from zero upon
crossing the transition, as opposed to exhibiting a finite jump.
This reflects the different nature of the two transition mechanisms:
while ordinary glass transitions are driven by collective caging
on local length scales, the localization transition contains a divergent
two-point length scale (the localization length of the small particles).
This also means that the applicability 
of MCT to such transitions can be debated (even more than MCT is usually
debated). The dashed line in Fig.~\ref{phicdiag}(a) is to be seen as
a good qualitative approximation to the localization line \cite{Schnyder},
as it is expected on physical grounds: increasing the density of the
single glass, one will eventually reach a regime where the small particles
either have too little void space to move, or are dense enough to form
a glass on their own.
For this reason, the single-glass region is bounded from below by the
liquid-glass transition, from above by a
localization transition, and bounded from the high-$\hat x$ side
by the glass--glass
transition in Fig.~\ref{phicdiag}(a). This is in variance with
Ref.~\cite{JuarezMaldonado2008}, where no glass--glass transition is found.
The separate localization transition in our calculation occurs ony below
some size ratio $\delta^s_c$ strictly smaller than the $\delta_c$ where
the higher-order singularity first occurs.
The precursor to the localization transition is anomalous diffusion
in the sense that the small-particle mean-squared displacement,
$\delta r^2(t)=\langle(\vec r_{s,\sm}(t)-\vec r_{s,\sm}(0))^2
\rangle$, exhibits power-law growth $\delta r^2(t)\propto t^y$ with an
exponent $y<1$ \cite{Sjoegren1986},
instead of ordinary diffusion ($y=1$), or the two-step
glass-transition pattern comprising an intermediate plateau ($y\approx0$).
This allows the existence of the ``double-transition'' scenario to be
established, as done recently for simulations of a soft-sphere mixture
\cite{wcamix}.

So far, we have found three distinct glasses separated by
well-defined transitions.
The glass--glass transition is similar to
the one found in colloid--polymer mixtures: both share the same
mathematics, and both arise from a competition of arrest mechanisms
on two sufficiently distinct length scales. Tempting as it may
be \cite{JuarezMaldonado2008b},
the analogy is flawed, since in the colloid--polymer mixture
the small component is always assumed to remain mobile, whereas at
the glass--glass transition in Fig.~\ref{phicdiag} the small component itself
becomes the main glass former.

\begin{figure}
\includegraphics[width=.9\linewidth]{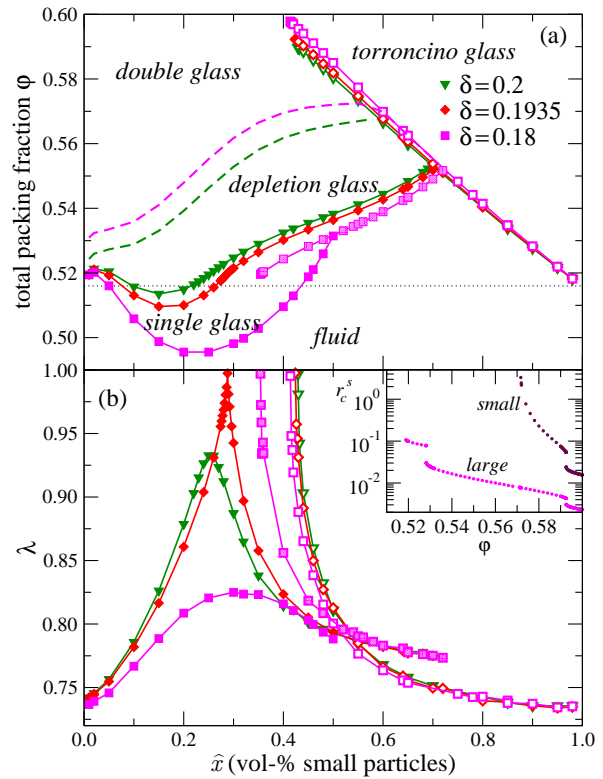}
\caption{\label{phicdiag2}
  Transition diagram as in Fig.~\ref{phicdiag}, but for smaller $\delta$.
  Open, filled, and shaded symbols correspond to
  $\lambda(\hat x)$ for the small-$\hat x$, large-$\hat x$, and
  intermediate-$\hat x$ branch of $\phi^c(\hat x)$, respectively.
  Inset: localization length $r^s_c$ for large and small particles
  for $(\delta,\hat x)=(0.18,0.45)$ as a function of packing fraction.
}
\end{figure}

The analog of the attractive glass in the binary mixture has to be
found inside the single glass region. Indeed, extending Fig.~\ref{phicdiag}
to smaller $\delta$, Fig.~\ref{phicdiag2},
a second glass--glass transition emerges: a second ${\mathcal A}_4$
point at roughly $\delta^*=0.194$ marks the onset of the higher-order
singularity scenario for $\delta<\delta^*$ at around $\hat x\approx0.3$.
Again, it indicates the discontinuous change of a big-particle dominated
glass to one where the small particles set a relevant length scale,
but now in the sense that
the small particles induce a strong depletion attraction while themselves
remaining mobile.

The different glasses are distinguished by the localization lengths
$r^s_{c,\alpha}$ of big and small particles -- a measure of their cage
size. The inset of
Fig.~\ref{phicdiag2}(b) shows $r^s_{c,\alpha}(\varphi)$ along a cut
crossing all four transitions.
As the repulsive single-glass is first entered,
$r^s_{c,\la}\approx0.1$, the Lindemann length for
big particles. It discontinuously drops by about $\delta$
entering the attractive single-glass, as depletion forces move the large
particles closer together. The small-particle
$r^s_{c,\sm}$ remains infinite up to the localization transition,
and shows signs of a continuous divergence there
(where $r^s_{c,\la}$ is smooth).
As the ``torroncino'' glass is entered, both $r^s_{c,\alpha}$ drop
discontinuously again,
and $r^s_{c,\sm}\approx0.1\delta$ indicates a small-particle
glass.


To summarize, four different glasses are predicted to form in
the simplest model of glass-forming mixtures, viz., binary
hard-sphere mixtures,
if both composition and size ratio are changed. The four glasses
come in two categories, one (``double glass'') where both species freeze
simultaneously, and one (``single glass'') where the smaller component
remains mobile inside the frozen environment.
Each of these comes
in two variants, ``repulsive'' and ``attractive'',
depending on whether the structure is primarily driven by large-particle
caging, or by the small-particle induced
forces (be they arrested or not). The attractive double glass naturally
explains low-big-particle coordination ``asymmetric'' cages thought to
arise from ultra-soft potentials.
Our calculations make it plausible to
expect the existence of four glasses much more generically, as long
as the relevant mixture parameters cover a sufficiently wide range of states
(which may, for example, also depend on dimensionality \cite{Hajnal2009}).

A unique prediction is that the four types of glass are
separated by sharp transitions and
accompanied by regions in parameter space where signatures of
higher-order singularities or of anomalous diffusion may be seen. The latter
was found in recent simulations \cite{wcamix}. Features of
logarithmic decay that likely are the signature of the first set of
${\mathcal A_3}$ singularities have been reported \cite{Moreno2006b}.
The possibility of attractive glasses emerging around
$\delta\approx0.2\approx\delta^*$ has been hinted at
\cite{Germain2009}.
Further simulations and experiments are called for to conclusively
test our predictions.

Cautionary remarks should be added:
first, the PY approximation has known defects.
Still it appears to predict $\boldsymbol S(q)$ surprisingly sensibly
\cite{Emanuela}.
The effects discussed above arise from an
interplay of different length scales; this should be
captured qualitatively correctly in PY.
Second, experiment and simulation may require to resort to more
complicated interaction potentials, to suppress equilibrium phase
transitions that could otherwise interfere.
Finally, MCT is driven to a regime where it may likely fail;
our predictions
thus pose a demanding test case for the theory.

\begin{acknowledgments}
This work was funded by the Helmholtz-Gemeinschaft (HGF), Young Investigator
Group VH-NG~406, and the Zukunftskolleg, Universit\"at Konstanz.
I thank Emanuela Zaccarelli for insightful cooperation and
acknowledge discussion with J\"urgen Horbach and Simon Schnyder.
\end{acknowledgments}

\bibliography{lit,litnotes}
\bibliographystyle{apsrev}

\end{document}